\documentclass[prb,twocolumn,showpacs,superscriptaddress,preprintnumbers,amssymb]{revtex4}
\usepackage{graphicx}% Include figure files
\usepackage{dcolumn}% Align table columns on decimal point
\usepackage{bm}% bold math
\usepackage{wasysym}
\newcommand{\beq}{\begin{equation}}
\newcommand{\eeq}{\end{equation}}
\newcommand{\beqn}{\begin{eqnarray}}
\newcommand{\eeqn}{\end{eqnarray}}

\begin{document}

\title{Exotic continuous quantum phase transition between $Z_2$ topological
spin liquid and N\'{e}el order}

\author{Eun-Gook Moon}

\author{Cenke Xu}

\affiliation{Department of Physics, University of California,
Santa Barbara, CA 93106}

\begin{abstract}

Recent numerical simulations with different techniques have all
suggested the existence of a continuous quantum phase transition
between the $Z_2$ topological spin liquid phase and a conventional
N\'{e}el order. Motivated by these numerical progresses, we
propose a candidate theory for such $Z_2 -$N\'{e}el transition. We
first argue on general grounds that, for a SU(2) invariant system,
this transition {\it cannot} be interpreted as the condensation of
spinons in the $Z_2$ spin liquid phase. Then we propose that such
$Z_2 - $N\'{e}el transition is driven by proliferating the bound
state of the bosonic spinon and vison excitation of the $Z_2$ spin
liquid, $i.e.$ the so called $(e, m)-$type excitation. Universal
critical exponents associated with this exotic transition are
computed using $1/N$ expansion. This theory predicts that at the
$Z_2 - $N\'{e}el transition, there is an emergent quasi long range
power law correlation of columnar valence bond solid order
parameter.
%, which is consistent with recent numerical results.

\end{abstract}

\date{\today}

\maketitle

\section{Introduction}

Thanks to the rapid development of numerical techniques, more and
more candidates of exotic liquid states have been identified in
frustrated spin
models~\cite{white,jiang2008,hongchen,gu2011,fabio}, hard-core
quantum boson model~\cite{kimz2,kimz22}, or even Hubbard
model~\cite{meng}. All these phases that are identified
numerically are fully gapped liquid phases with short range
correlation between both spin order parameters and also valence
bond solid (VBS) order parameters. The simplest fully gapped spin
liquid state is the $Z_2$ topological liquid state, which has the
same topological order as the toric code model~\cite{kitaev1997}.
In addition to the fully gapped spectrum, the computation of
critical exponents at the order-disorder transition of these
models~\cite{kimz2,kimz22,melko3}, and the computation of
topological entanglement entropy~\cite{hongchen,melko2} both
convincingly proved that the spin liquid states of some of these
models (such as the $J_1 - J_2$ model on the square lattice, and
the extended Bose-Hubbard model on the Kagome lattice) are indeed
the $Z_2$ topological liquid. In some other models (such as the
spin-1/2 Heisenberg model on the Kagome
lattice~\cite{white,jiang2008}, and the Hubbard model on the
honeycomb lattice~\cite{meng}), although an accurate topological
entanglement entropy computation is still demanded, it is broadly
believed that the spin liquid state is indeed the $Z_2$ liquid
state, or a similar $(Z_2)^n $ liquid state.

Besides the spin liquid state itself, the quantum phase
transitions of these models are equally interesting. For example,
continuous quantum phase transitions between N\'{e}el order and a
fully gapped spin liquid phase have been found in the honeycomb
lattice Hubbard model~\cite{meng}, and the $J_1-J_2$ spin-1/2
Heisenberg model on the square lattice~\cite{hongchen,gu2011}. %EM1 subject : phases (plural)
In terms of the Landau-Ginzburg (LG) theory, this transition
should be an ordinary O(3) transition, and the $Z_2$ liquid phase
is identified as the disordered phase, while the N\'{e}el phase is
the ordered phase. However, because the $Z_2$ liquid phase has a
nontrivial topological order and topological
degeneracy~\cite{kitaev1997}, it cannot be adiabatically connected
to the trivial direct product state, thus it should {\it not} be
identified as the trivial disordered phase in the classical case.
Thus if the $Z_2-$N\'{e}el transition exists, it means that the
quantum disordering of the N\'{e}el order and the emergence of the
$Z_2$ topological order happen simultaneously at one point, this
unusual fact implies that this $Z_2-$N\'{e}el quantum critical
point (QCP) must be an unconventional one that is beyond the LG
paradigm. The goal of this paper is to understand this
unconventional QCP.

\section{Failure of the spinon theories}

We first argue on general grounds that such a continuous
$Z_2-$N\'{e}el transition {\it cannot} be understood using an
ordinary spinon theory. We stress that we will only consider SU(2)
invariant systems here.

First of all, if this $Z_2$ spin liquid phase has a gapped
fermionic spinon excitation $f_\alpha$, then a N\'{e}el order
parameter can in principle be represented as $\vec{N} \sim (-1)^i
f^\dagger_{i, \alpha} \vec{\sigma}_{\alpha \beta} f_{i, \beta}$.
Thus it appears that we can interpret this $Z_2-$N\'{e}el
transition as the disorder-order transition of the vector $\vec{N}$
using an ordinary Landau-Ginzburg theory. However, this theory is
incorrect because the vector $\vec{N}$  does not carry any
gauge charge, thus the order parameter does not immediately
suppress the $Z_2$ topological order. This implies that between
the $Z_2$ spin liquid and the N\'{e}el order with nonzero $\langle
\vec{N} \rangle$, there must be an intermediate state with the
coexistence of both N\'{e}el order and $Z_2$ topological order,
and it is usually called the N\'{e}el$^\ast$ state. Thus a direct
continuous transition between $Z_2$ and N\'{e}el order cannot be
obtained this way without fine-tuning.

In order to suppress the $Z_2$ topological order, the usual wisdom
is to condense a topological excitation that carries the $Z_2$
gauge charge. Then after the topological excitations are
condensed, the $Z_2$ gauge field is Higgsed, and the topological
order disappears. Along with suppressing the topological order, if
we want to induce spin order simultaneously, then the excitation
that condenses
% EM3 the condensed excitation
must also carry certain representation of the spin SU(2) symmetry
group, in addition to the $Z_2$ gauge charge. Let us call this
gauge-charged spin excitation as {\it spinon} in general. Then the
nature of the spin order and the universality class of this
transition both depend on the particular spin representation of
spinon.

The smallest representation of SU(2) is spin-1/2 representation,
and there is no consistent ``fractional" representation of SU(2)
group that is smaller than spin-1/2. Thus let us first assume the
spinon is a spin-1/2 boson, which is described by a two component
complex boson field $z_\alpha = (z_1, z_2)^t$, and $z_\alpha$ is
subject to the constraint $|z_1|^2 + |z_2|^2 = 1$. Then $z_\alpha$
is coupled to a $Z_2$ gauge field in the following way: \beqn H =
\sum_{i, \mu}\sum_\alpha - t \sigma^z_{i, \mu} z^\ast_{\alpha, i}
z_{\alpha, i+\mu} + H.c. + \cdots \label{o4field} \eeqn where the
ellipsis stands for higher order interaction terms.
$\sigma^z_{i,\mu}$ is the $Z_2$ gauge field that is defined on the
link $(i, \mu)$ of the lattice, and Eq.~\ref{o4field} is invariant
under the gauge transformation \beqn z_{i,\alpha} \rightarrow
\eta_i z_{i,\alpha}, \ \ \ \sigma^z_{i, \mu} \rightarrow \eta_{i}
\sigma^z_{i, \mu} \eta_{i+\mu}, \eeqn where $\eta_i = \pm 1$ is an
arbitrary Ising function defined on the sites of the lattice.
 The condensed phase of $z_\alpha$ is the spin ordered phase, and
because $z_\alpha$ is coupled to the $Z_2$ gauge field, the $Z_2$
topological order is automatically destroyed due to the Higgs
mechanism in the condensate of $z_\alpha$. The gapped phase of
$z_\alpha$ is the deconfined $Z_2$ topological phase.

Since $z_\alpha$ has in total two complex bosonic fields, $i.e.$
four real fields, then with the constraint $|z_1|^2 + |z_2|^2 =
1$, the entire configuration of $z_\alpha$ is equivalent to a
three dimensional sphere $S^3$. Since the spinon field $z_\alpha$
is coupled to a $Z_2$ gauge field, then the physical configuration
of the condensate of $z_\alpha$ is $S^3/Z_2$, which is
mathematically equivalent to the group manifold SO(3). Since
$z_\alpha$ itself is not a physical observable, inside the
condensate of $z_\alpha$ the physical observables are the three
following vectors: \beqn \vec{N}_1 = \mathrm{Re}[z^t i\sigma^y
\vec{\sigma} z], \ \ \vec{N}_2 = \mathrm{Im}[z^t i\sigma^y
\vec{\sigma} z], \ \ \vec{N}_3 = z^\dagger \vec{\sigma} z. \eeqn A
simple application of the Fierz identity $\sum_a
\sigma^a_{\alpha\beta} \sigma^a_{\gamma\rho} =
2\delta_{\alpha\rho}\delta_{\beta\gamma} -
\delta_{\alpha\beta}\delta_{\gamma\rho}$ proves that these three
vectors are orthogonal with each other. Since the first homotopy
group of SO(3) is $\pi_1[\mathrm{SO(3)}] = Z_2$, inside this spin
ordered phase there are vortex-like topological defects. Two of
these vortices can annihilate each other.

The spin-1/2 boson field $z_\alpha$ can be viewed as the low
energy mode of the usual Schwingber boson $b_\alpha$, but our
argument is more general, and it is independent of the microscopic
origin of $z_\alpha$. If we identify one of the three vectors
$\vec{N}_i$ as the N\'{e}el vector, then this phase must have two
other spin vector orders that are perpendicular to the N\'{e}el
vector. The condensation transition of $z_\alpha$ while coupled to
a $Z_2$ gauge field is usually called the O(4)$^\ast$
transition~\cite{senthilchubukov}.

%Clearly when the spinon carries a spin-1/2 representation, the
%spin order of the spinon condensate is different from the N\'{e}el
%order.

Now let us assume the spinon of the $Z_2$ topological phase
carries a spin-1 representation. A spin-1 representation is a
vector representation of SU(2), $i.e.$ it can be parametrized as a
unit real vector $\vec{n}$, $ |\vec{n}|^2 = 1$. Now the coupling
between the spinon and $Z_2$ gauge theory reads \beqn H = \sum_{i,
\mu}\sum_a - t \sigma^z_{i, \mu} n^a_{i} n^a_{i+\mu}  + \cdots
\eeqn Again, since $\vec{n}$ couples to a $Z_2 $ gauge field, it
is not a physical observable: $\vec{n}$ and $-\vec{n}$ are
physically equivalent. If vector $\vec{n}$ condenses, the
condensate is in fact a spin nematic, or quadrupole order, with
physical order parameter \beqn Q^{ab} = n^a n^b - \frac{1}{3}
\delta_{ab} . \eeqn This spin order has manifold $S^2/ Z_2$, which
also supports vortex excitation since $\pi_1[S^2/Z_2] = Z_2$. One
example state of this type is the spin quadrupolar state that has
been observed in the spin-1 material
$\mathrm{NiGa_2S_4}$~\cite{ngs}.

%The condensation transition of the vector $\vec{n}$ belongs to the
%3D O(3)$^\ast$ universality class.

We have discussed two types of unconventional QCPs between $Z_2$
liquid phase and spin orders. In either case, the spin ordered
phase is different from the ordinary collinear N\'{e}el order,
because a N\'{e}el order should have ground state manifold (GSM)
$S^2$. In particular, in both cases we have considered, the spin
ordered phase must have a nontrivial homotopy group $\pi_1$, which
corresponds to the vison excitation of the $Z_2$ gauge field.
Generalization of our analysis to higher spin representations is
straightforward, but the conclusion is unchanged. As we already
discussed, in Ref.~\cite{hongchen} and Ref.~\cite{meng}, a {\it
continuous} quantum phase transition between a fully gapped spin
liquid phase and a N\'{e}el order was reported. If the fully
gapped spin liquid discovered in these numerical works is indeed a
$Z_2$ spin liquid as we expected, then such continuous quantum
phase transition is beyond the spinon theory discussed in this
section. In order to understand the continuous transition between
the gapped spin liquid and N\'{e}el order reported in the phase
diagram of the Hubbard model on the honeycomb lattice \cite{meng},
in Ref.~\cite{wang,ran,xu2010} the authors had to introduce extra
``hidden" order parameters in the N\'{e}el phase, which change the
GSM of the N\'{e}el phase completely.

In this section we argued on general grounds that the
$Z_2-$N\'{e}el transition cannot be interpreted as the
condensation of an ordinary spinon. Our argument is independent of
specific spin model or lattice structure. However, this argument
can only be applied to SU(2) invariant systems. For a system with
U(1) symmetry, for example the hard-core Boson model on the Kagome
lattice discussed in Ref.~\cite{balentsz2,melko1,melko2,melko3},
the transition between $Z_2$ topological phase and the superfluid
phase can be understood as the condensation of a fractionalized
``half-boson" that couples to the $Z_2$ gauge field, and this
transition is the so-called $3d$ XY$^\ast$ transition.
%EM5

\section{Exotic Z$_2-$N\'{e}el Quantum critical point}

\subsection{Phase diagram around $Z_2$ spin liquid driven by $e$ and $m$ excitations}

In order to understand the direct continuous transition between
the $Z_2$ spin liquid and the N\'{e}el phase, we should first put
these two phases in the same phase diagram. One candidate theory
that contains both phases was proposed in Ref.~\cite{xusachdev}.
Let us first write down a minimal unified field theory proposed in
Ref.~\cite{xusachdev}: \beqn \mathcal{L} &=& \sum_{\alpha =
1}^{N_z} |(\partial_\mu - i a_\mu) z_\alpha|^2 + \sum_{\alpha =
1}^{N_v} |(\partial_\mu - i b_\mu) v_\alpha|^2 \cr\cr &+& s_z
|z_\alpha|^2 + s_v |v_\alpha|^2  +
\frac{i}{\pi}\epsilon_{\mu\nu\rho} a_\mu
\partial_\nu b_\rho + \cdots \label{mcs}\eeqn In this field theory, there
are two types of matter fields, $z_\alpha$ and $v_\alpha$, and
they are interacting with each other through a mutual Chern-Simons
(CS) theory, which grants them a mutual semionic statistics $i.e.$
when $v_\alpha$ adiabatically encircles $z_\alpha$ through a
closed loop, the system wave-function acquires a minus sign. This
is one of the key properties of the $Z_2$ topological phase. Here
$z_\alpha$ corresponds to the electric ($e-$type) excitation of
the $Z_2$ liquid, and $v_\alpha$ corresponds to the magnetic
($m-$type) excitation. $v_\alpha$ is usually called the vison
excitation.

The minimal field theory Eq.~\ref{mcs} has symmetry
SU($N_z$)$\times$SU($N_v$). However, depending on the details of
the microscopic model, the higher order interactions between
matter fields can break this symmetry down to its subgroups. We
will first ignore this higher order symmetry breaking effects, and
focus on the case with $N_z = 2$, and $N_v = 1$. In
Ref.~\cite{xusachdev}, the authors used the model Eq.~\ref{mcs}
with $N_z =2$, $N_v = 1$ to describe the global phase diagram of
spin-1/2 quantum magnets on a distorted triangular lattice, which
is a very common structure in many materials. The same theory can
be applied to the square and honeycomb lattice as well, and in
this paper we will take the square lattice as an example. Here
$z_\alpha$ is a bosonic spin-1/2 spinon, and $v$ is the low energy
mode of a vison, and it corresponds to the expansion of the vison
at two opposite momenta $\pm \vec{Q} $: \beqn \tau \sim v
e^{i\vec{Q}\cdot \vec{r}} + v^\ast e^{- i\vec{Q}\cdot \vec{r}},
\eeqn thus $v$ is a complex scalar field. On the square lattice or
distorted triangular lattice, there is a $Z_8$ anisotropy on $v$,
that is allowed by the symmetry of the lattice
\cite{xusachdev,xubalents}. This anisotropy is highly irrelevant
in the quantum critical region, and it will be ignore throughout
the paper.

The phase diagram of this model is tuned by two parameters: $s_z$
and $s_v$, and depending on the sign of these two parameters,
there are in total four different phases (Fig.~\ref{QCP}):

\begin{figure}
\begin{center} \includegraphics[width=3.1 in]{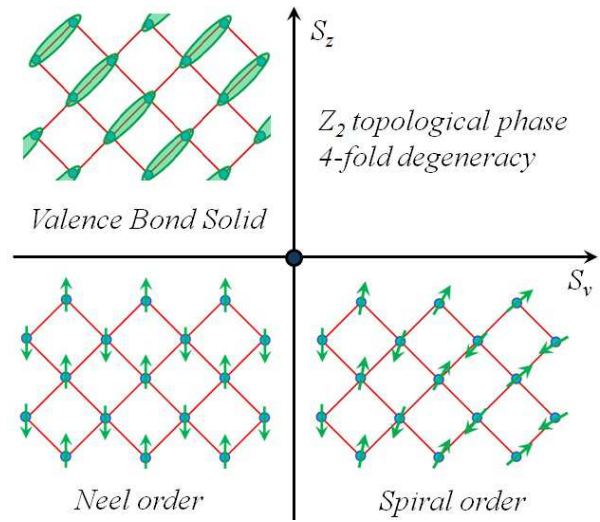}
\end{center}
\caption{The global phase diagram of Eq.~\ref{mcs}, which
describes four different states on a distorted triangular lattice,
or a square lattice. Eq.~\ref{mcs} assumes that the $e-$type
excitation $z_\alpha$ and $m-$type excitation $v$ condense
separately.} \label{QCP}
\end{figure}

{\it Phase 1.} This is the phase with $s_z > 0$, $s_v > 0$. In
this phase, both matter fields $z_\alpha$ and $v$ are gapped, and
they have a topological statistic interaction through the mutual
CS theory. Since all the matter fields are gapped, the low energy
properties of phase 1 is described by the mutual CS theory only.
The mutual CS theory defined on a torus has a four-fold degenerate
ground state, thus this phase is precisely the gapped $Z_2$
topological phase~\cite{kou}.

{\it Phase 2.} $s_v > 0$, $s_z < 0$. When vison $v$ is gapped,
integrating out $v$ induces a Maxwell term for gauge field
$b_\mu$, which implies that the flux of $b_\mu$ is condensed. In
other words the flux-creation operator (denoted as
$\mathcal{M}_b$) acquires a nonzero expectation value.
$\mathcal{M}_b$ corresponds to a Dirac monopole configuration of
$b_\mu$ in the space-time. Due to the mutual CS coupling between
gauge fields $a_\mu$ and $b_\mu$, the condensate of
$\mathcal{M}_b$ breaks $a_\mu$ to a $Z_2$ gauge field. Thus after
we integrate out $v$ and $b_\mu$, the spinon $z_\alpha$ is only
coupled to a $Z_2$ gauge fields. Thus when $N_z = 2$, the
condensate of $z_\alpha$ has GSM SO(3) as was discussed in the
previous section. An example of this phase is the spiral spin
density wave phase. Once we assume $s_v
> 0$, Eq.~\ref{mcs} precisely reduces to the
previously studied O(4)$^\ast$ theory for the transition between
$Z_2$ spin liquid and spiral spin order~\cite{senthilchubukov}.

{\it Phase 3.} $s_v < 0$, $s_z > 0$. This is a phase where $v$
condenses while $z_\alpha$ is gapped out. This phase is the four
fold degenerate columnar VBS phase that breaks the reflection and
translation symmetry of the lattice. The columnar VBS order
parameter can be written as $v^2 \mathcal{M}_a$, where
$\mathcal{M}_a$ is the monopole operator of gauge field $a_\mu$,
which creates a $2\pi$ flux of $a_\mu$. When $s_z > 0$, spinon
$z_\alpha$ is gapped, and it leads to a Maxwell term for $a_\mu$.
This implies that $\mathcal{M}_a$ is condensed, and it breaks
$b_\mu$ to a $Z_2$ gauge field. In this case the low energy
effective theory that describes phase 3 is a complex field $v$
that couples to a $Z_2$ gauge field, thus
%$v^2
%\mathcal{M}_a$ is equivalent to $v^2$, and
our theory reduces to the pure vison theory that was thoroughly
discussed in Ref.~\cite{xubalents}.

{\it Phase 4.} $s_v < 0$, $s_z < 0$. This is a phase where both
$z_\alpha$ and $v$ condense. Because in this phase the only gauge
invariant order parameter that condenses is $\vec{N} \sim
z^\dagger \vec{ \sigma} z$, this is precisely the collinear
N\'{e}el phase with GSM $S^2$. In fact, when $v$ is condensed, the
gauge field $b_\mu$ acquires a mass term $b_\mu^2$ due to the
Higgs mechanism. Then integrating out $v$ and $b_\mu$ leads to a
Maxwell term for gauge field $a_\mu$, due to the mutual CS
coupling. Thus the spinon $z_\alpha$ is coupled to a dynamical
gapless U(1) gauge field $a_\mu$. Then the GSM of the condensate
of $z_\alpha$ is $S^3/U(1) = S^2$, which is equivalent to the
collinear N\'{e}el order. Thus under the assumption $s_v < 0$,
Eq.~\ref{mcs} reduces to the CP(1) model that describes the
deconfined QCP between N\'{e}el and VBS
order~\cite{deconfine1,deconfine2}.

We have shown that the mutual CS formalism Eq.~\ref{mcs} unifies
many previously discussed exotic states and exotic phase
transitions. A more detailed discussion of the phase diagram can
be found in Ref.~\cite{xusachdev}.

%The QCP between phase 1 and 2 is the 3D O(4)$^\ast$ transition;
%The QCP between phase 1 and 3 is the 3D XY$^\ast$ transition. The
%QCP between phase 3 and 4 is the deconfined QCP that is described
%by the noncompact CP(1) field theory.

\subsection{ $Z_2-$N\'{e}el transition driven by $(e, m)$ excitation }

Now we are ready to discuss our theory for the direct continuous
transition between $Z_2$ liquid phase and N\'{e}el phase. In a
$Z_2$ topological phase, using the standard notation, there are
three types of topological excitations: the electric excitation
$e$, the magnetic excitation $m$, and their bound state $(e, m)$.
In Eq.~\ref{mcs}, the spinon field $z_\alpha$ is the $e-$type
excitation, while the vison field $v$ is the $m-$type excitation.
Eq.~\ref{mcs} is based on the assumption that inside the $Z_2$
liquid phase the $e-$type and $m-$type excitations have lower
energy than $(e, m)$, thus in the global phase diagram
Fig.~\ref{QCP}, the N\'{e}el and $Z_2$ topological phases are
separated by a multicritical point $s_z = s_v = 0$. However, if we
consider the opposite possibility, namely the $(e, m)-$type
excitation has the lowest energy in the $Z_2$ spin liquid, then a
different quantum phase transition can occur by condensing the
$(e, m)-$type excitation.
%EM7
%It is expected that a finite local attractive interaction between
%$e$ and $m$ excitations would lead to a bound state $(e, m)$ with
%lower excitation energy in the $Z_2$ liquid phase.

\begin{figure}
\begin{center} \includegraphics[width=3.0 in]{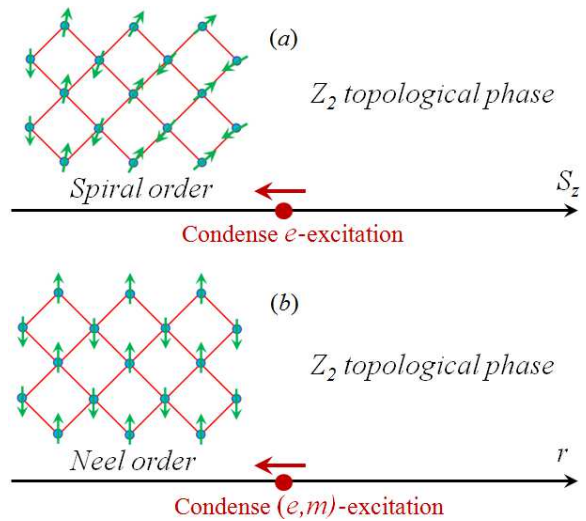}
\end{center}
\caption{($a$) For a SU(2) invariant system, if one condenses the
ordinary spinon of the $Z_2$ spin liquid phase, the spin ordered
state must have ground state manifold SO(3). One example state of
this kind is the spiral spin density wave. ($b$) If the $(e, m)$
type of excitation of the $Z_2$ spin liquid condenses, the spin
order can be the ordinary N\'{e}el order. At the $Z_2-$N\'{e}el
QCP, both N\'{e}el and columnar VBS order parameters have
power-law correlation. } \label{QCP2}
\end{figure}

Let us first take the simplest Toric code model~\cite{kitaev1997}
as an example: $H = \sum_i - \sigma^x_{i, - x }\sigma^x_{i,
x}\sigma^x_{i, - y}\sigma^x_{i, y} - \sigma^z_{i, x }\sigma^z_{i,
y}\sigma^z_{i + x, y}\sigma^z_{i + y, x}$. The ``condensation" of
an excitation simply means that the system enters a phase were the
kinetic energy of this excitation dominates. It is well-known that
in the Toric code model the condensation of the $e-$excitation is
driven by a magnetic field $h_z \sigma^z_{i,\mu}$, while the
condensation of $m-$excitation is driven by field $h_x
\sigma^x_{i,\mu}$, because these two fields enable the hopping of
$e$ and $m$ excitations respectively. In order to ``condense" the
$(e,m)$ excitation, we simply need to turn on field $h_y
\sigma^y_{i,\mu}$, which hops the $(e,m)$ excitation along the
diagonal directions of the square lattice. When any of the three
excitations is condensed, the system enters a trivial polarized
state without any topological degeneracy. Generally speaking, in
the topological phase, starting from one of the topological
sectors on the torus, the other sectors can be generated by
locally creating a pair of topological excitations, and
annihilating them after adiabatically moving one excitation of the
pair around the torus. Because all three types of topological
excitations are mutual semions, condensing one of the three
excitations will lead to a strong local flux fluctuation for the
other two excitations, thus the other two excitations are confined
in this condensate, $i.e.$ the system no longer has topological
degeneracy.

In our current case, both $e$ and $m$ excitations carry extra
global symmetries besides their gauge charges. In order to
describe the $(e,m)$ excitation in our situation, let us define
new complex bosonic fields $\phi_\alpha$ and $\psi_{\alpha}$:
\beqn \phi_\alpha = z_\alpha v, \ \ \ \ \psi_\alpha = z_\alpha
v^\ast. \eeqn $\phi_\alpha$ and $\psi_\alpha$ carry the quantum
number of $(e, m)$ excitation. Because $v$ is a complex variable,
fields $\phi_\alpha$ and $\psi_\alpha$ are independent from each
other, and they interact with each other as follows: \beqn
\mathcal{L} &=& \sum_{\alpha} |(\partial_\mu - i a_\mu - i b_\mu )
\phi_\alpha|^2 + |(\partial_\mu - i a_\mu + i b_\mu )
\psi_\alpha|^2 \cr\cr &+& r (|\phi_\alpha|^2 + |\psi_\alpha|^2 ) +
\frac{i}{\pi} \epsilon_{\mu\nu\rho} a_\mu
\partial_\nu b_\rho \cr\cr &+& g (|\phi|^2)^2 + g (|\psi|^2)^2 + u |\phi|^2|\psi|^2
- w \phi^{\dagger} \vec{\sigma} \phi \cdot \psi^{\dagger}
\vec{\sigma} \psi. \label{ab}\eeqn Notice that $\phi_\alpha$ and
$\psi_\alpha$ carry gauge charges of both gauge fields $a_\mu$ and
$b_\mu$. In order to understand the QCP at $r = 0$ more
quantitatively, it is more convenient to define new gauge field
$A^{\pm}_\mu = a_\mu \pm b_\mu$, then the Lagrangian reads
%\footnote{Rigorously speaking, $a_\mu$ and $b_\mu$ are both
%compact U(1) gauge fields, but in order to understand the quantum
%critical point at $r =0 $ in Eq.~\ref{ab}, we assume that their
%space-time instantons are irrelevant at the QCP, $i.e.$ $a_\mu$
%and $b_\mu$ become noncompact at the QCP. This assumption is
%justified when the number of the matter fields component $N$ is
%large enough, since it is well-known that the scaling dimension of
%instanton is proportional to $N$ when the matter field is gapless.
%Under this assumption, $A^+_\mu$ and $A^-_\mu$ can also be treated
%as noncompact fields.}
: \beqn \mathcal{L} &=& \sum_{\alpha}
|(\partial_\mu - i A^{+}_\mu ) \phi_\alpha|^2 + |(\partial_\mu - i
A^-_\mu ) \psi_\alpha|^2 \cr\cr &+& r (|\phi_\alpha|^2 +
|\psi_\alpha|^2 ) + \frac{i}{4\pi} \epsilon_{\mu\nu\rho} A^+_\mu
\partial_\nu A^+_\rho - \frac{i}{4\pi} \epsilon_{\mu\nu\rho}
A^-_\mu \partial_\nu A^-_\rho \cr\cr &+& g (|\phi|^2)^2 + g
(|\psi|^2)^2 + u |\phi|^2|\psi|^2 - w \phi^{\dagger} \vec{\sigma}
\phi \cdot \psi^{\dagger} \vec{\sigma} \psi. \label{b1b2} \eeqn

In this field theory, $\phi_\alpha$ and $\psi_\alpha$ are almost
decoupled from each other, $i.e.$ they are only coupled through
the quartic terms $u$ and $w$. The mass gaps $r$ for $\phi_\alpha$
and $\psi_\alpha$ are equal, because the vison modes $v$ and
$v^\ast$ are guaranteed to be degenerate by the symmetry of the
square lattice. $\phi_\alpha$ and $\psi_\alpha$ are introduced as
bosonic fields, but gauge fields $A^+_\mu$ and $A^-_\mu$ make them
fermionic fields after the standard flux attachment, due to the
existence of the Chern-Simons terms in this Lagrangian. In our
formulation, fields $\phi_\alpha$ and $\psi_\alpha$ can still
condense by tuning parameter $r$ in Eq.~\ref{b1b2}. After
$\phi_\alpha$ and $\psi_\alpha$ both condense simultaneously,
$A^+_\mu$ and $A^-_\mu$ are both Higgsed, and in the Higgs phase
the only gauge invariant operators are \beqn \phi^{\dagger}
\vec{\sigma} \phi, \ \ \ \ \psi^{\dagger} \vec{\sigma} \psi. \eeqn
Since these two vectors both carry the same quantum number as the
N\'{e}el order parameter $z^\dagger \vec{\sigma} z$, in
Eq.~\ref{b1b2} $w$ is naturally positive, thus these two vectors
are aligned parallel with each other, so the condensate of
$\phi_\alpha$ and $\psi_\alpha$ has a manifold $S^2$, $i.e.$ it is
the standard collinear N\'{e}el order. The difference between this
new transition and the ordinary spinon theory is illustrated in
Fig.~\ref{QCP2}.

What is the universality class of this transition? The simplest
possibility is that, both $u$ and $w$ are irrelevant at the
transition, although they are relevant in the condensate of
$\phi_\alpha$ and $\psi_\alpha$. Under this assumption
$\phi_\alpha$ and $\psi_\alpha$ are completely decoupled at the
transition $r = 0$. Then in this case this transition is described
by the simple Chern-Simons-Higgs model: \beqn \mathcal{L} &=&
\sum_{\alpha = 1}^N |(\partial_\mu - i A_\mu ) \phi_\alpha|^2  + r
|\phi_\alpha|^2  + g (\sum_\beta |\phi_\beta|^2)^2 \cr\cr &+&
\frac{i N}{8 \theta} \epsilon_{\mu\nu\rho}A_\mu\partial_\nu
A_\rho. \label{ntheta}\eeqn Here we have generalized the equation
to have $N$ flavors of matter fields $\phi_\alpha$, and introduced
a statistical angle $\theta$. Our physical situation corresponds
to $N = 2$ and $\theta = \pi$. Notice that Eq.~\ref{ntheta}
explicitly breaks the time-reversal symmetry due to the
Chern-Simons term. But the complete theory Eq.~\ref{b1b2} is
time-reversal invariant, because under time-reversal
transformation $\phi_\alpha$ and $\psi_\alpha$ are exchanged, the
two gauge fields $A^+_\mu$ and $A^-_\mu$ are also exchanged.

The critical exponents of this transition can be computed using a
systematic $1/N$ expansion. Ref.~\cite{wen} has computed the
critical exponent $\nu$ defined as $\xi \sim |r|^{-\nu}$, here we
will focus on the scaling dimension of $\phi^\dagger T^a \phi$ at
the QCP, where $T^a$ is the SU(N) generator. To the first order
$1/N$ expansion, this scaling dimension reads \beqn
\Delta[\phi^\dagger T^a \phi] = 1 + \frac{4}{3\pi^2} \left(
\frac{4}{N} - \frac{1}{N} \frac{\theta^2/4}{1 + \theta^2/64}
\right). \eeqn Let us briefly comment how we obtain this result.
Similar calculation without the $\theta$ term was obtained before.
See Fig.3  and Fig.4 of the previous work\cite{sachdevcpn} for
necessary Feynman diagrams. First, we need to evaluate wave
function renormalization of $\phi$ from both gauge fluctuation and
the density fluctuation, which contain the factor($1/N$). Then,
using the standard operator insertion method, one can calculate
renoramlization function of the corresponding vertex. We note that
the traceless condition, (${\rm Tr}( T^a) =0$), reduces one
diagram compared with the calculation of the scaling dimension of
$|\vec{\phi}|^2$ and simplify our calculation.

In the limit of $\theta \rightarrow \infty$, $i.e.$ the CS term is
effectively zero, these results converges to the ordinary
CP($N-1$) results computed in Ref.~\cite{sachdevcpn}. In the limit
of $\theta \rightarrow 0$, the gauge fluctuation is totally frozen
by the CS term, and the universality class of this quantum
critical point only acquires corrections from the short range
self-interaction between field $\phi_\alpha$, thus it is
equivalent to an O($2N$) transition of the O($2N$) bosonic vector
field $(\mathrm{Re}[\phi_1],\cdots \mathrm{Re}[\phi_N],
\mathrm{Im}[\phi_1], \cdots \mathrm{Im}[\phi_N])$. Scaling
dimension of the ``N\'{e}el'' type operator ($\Delta[\phi^\dagger
T^a \phi] $) in our theory is larger than that in the CP($N-1$)
theory with large $N$, $i.e.$ at the $Z_2-$N\'{e}el QCP, the
anomalous dimension of the N\'{e}el order parameter is predicted
to be larger than that of the deconfined QCP between the N\'{e}el
and VBS order. This prediction can be tested in the future by a
careful comparison between the critical exponents of the $J_1 -
J_2$ model and the $J-Q$ model~\cite{sandvik1,ribhu,sandvikvbs}.

It is pretty clear that at least in the large$-N$ limit, the
perturbation of $u$ in Eq.~\ref{b1b2} is irrelevant, because in
this limit the scaling dimension $\Delta[|\vec{\phi}|^2] =
\Delta[|\vec{\psi}|^2] = 2$, $i.e.$ $\Delta[u] = -1$. Higher order
$1/N$ or $\epsilon$ expansion is demanded to determine whether $w$
is relevant or not at this transition.

Assuming at the QCP $r = 0$ both $u$ and $w$ are irrelevant, then
besides the N\'{e}el order parameter, some other physical order
parameters also have power-law correlation. For example, the
columnar VBS order parameter can be written as \beqn \mathrm{VBS}
\sim \psi_\alpha^\dagger \phi_\alpha \mathcal{M}_a \sim v^2
\mathcal{M}_a, \eeqn where $\mathcal{M}_a$ is the monopole
operator for gauge field $a_\mu$. When $\phi_\alpha$ and
$\psi_\alpha$ both have a large $N$ component, the scaling
dimension of $ \mathcal{M}_a $ is proportional with $N$. Thus with
large $N$ the VBS order parameter is expected to have a much
larger scaling dimension compared with the N\'{e}el order
parameter at the $Z_2-$N\'{e}el QCP. We stress that the VBS order
parameter has short-range correlation in the $Z_2$ spin liquid and
the N\'{e}el phase, its emergent quasi long range correlation
occurs {\it only} at the QCP. This result has already been
confirmed numerically in Ref.~\cite{gu2011}, and it was
demonstrated that the scaling dimension of the VBS order parameter
is indeed larger than that of N\'{e}el order at the
QCP~\cite{gu2011}.

In 2+1 dimension, the entanglement entropy of a conformal field
theory can in general be written as $ S = c L - \beta$, where the
first term is the nonuniversal area law contribution, while the
second term is a universal constant. In
Ref.~\cite{swinglesenthil}, it was argued that at a QCP where a
bosonic field condenses while coupling to a discrete gauge field,
the universal entanglement entropy is a direct sum of the
contribution from the bosonic matter field and the contribution
from the discrete gauge field: $\beta = \beta_b + \beta_{gauge}$.
This conclusion is based on the assumption that the matter field
dynamics is not affected by the discrete gauge field in the
infrared limit, and this is indeed true for the XY$^\ast$
transition observed in Ref.~\cite{melko3}. However, at the exotic
$Z_2-$N\'{e}el transition discussed here where the $(e, m)-$type
excitations condense, the bosonic matter fields $\phi_\alpha$ and
$\psi_\alpha$ are indeed strongly affected by the gauge field,
thus at this QCP the universal entanglement entropy $\beta$ is no
longer a direct sum of the two different degrees of freedom of the
system. The universal entanglement entropy of field theory
Eq.~\ref{ntheta} in the large$-N$ limit can be found in
Ref.~\cite{sachdeventropy}.

\subsection{A Toy model with $N  = 1$}

Now let us discuss a toy model with $N = 1$. This is actually the
case where the critical theories can be all understood exactly.
This field theory with $N = 1$ can be applied to the following
extended Toric-code model: \beqn H &=& \sum_{i}K_x \sigma^x_{i, -
x }\sigma^x_{i, x}\sigma^x_{i, - y}\sigma^x_{i, y} + K_z
\sigma^z_{i, x }\sigma^z_{i, y}\sigma^z_{i + x, y}\sigma^z_{i + y,
x} \cr\cr & + & \sum_{i, \mu} h_x \sigma^x_{i,\mu} + h_z
\sigma^{z}_{i,\mu} + \cdots. \eeqn Here the $e-$type ($m-$type)
excitation is the end of a string product of $\sigma^x$
($\sigma^z$). The $e$ and $m-$type excitations view $\sigma^z$ and
$\sigma^x$ as $Z_2$ gauge fields respectively, and the $h_x$ and
$h_z$ terms enable the hopping of these excitations. Unlike the
standard toric-code model \cite{kitaev1997}, here we keep $K_x, \
K_z > 0$. When $K_x, \ K_z
> 0$, both $\sigma^z$ and $\sigma^x$ have a $\pi-$flux in the
ground state. Then the dynamics of both $e$ and $m$ type of
excitations are frustrated, and both excitations have two
different minima $\pm \vec{Q}$ in their band structure. As a
result, the low energy dynamics of $e$ and $m$ excitations are
described by complex scalar fields $z$ and $v$ expanded at
momentum $\vec{Q}$. If one of these two fields condenses while the
other one remains gapped, the $Z_2$ topological order is
destroyed, and the system must spontaneously break the lattice
translation symmetry as well. The condensates of $e$ and $m-$type
excitations physically corresponds to the valence bond solid phase
of $\sigma^z$ and $\sigma^x$ respectively.

When $N = 1$, if $e$ and $m$ type of excitations condense
separately, then the phases in Fig.~\ref{QCP} would be $Z_2$
liquid, VBS order of $\sigma^x$, trivial phase, and VBS order of
$\sigma^z$ (counted counterclockwise around the multicritical
point $s_z = s_v = 0$). On the other hand, if the bound state $(e,
m)$ has the lowest energy in the $Z_2$ liquid phase, then again we
can introduce two independent complex fields $\phi$ and $\psi$ as
$\phi = z v$, $\psi = z v^\ast$. Then the transition driven by the
condensation of $\phi$ and $\psi$ is described by Eq.~\ref{ntheta}
with $N = 1$ and $\theta = \pi/2$.

What kind of transition is this? If in Eq.~\ref{ntheta} the
complex field $\phi$ is also coupled to an external U(1) gauge
field $A^{ext}_\mu$, then we can see that in the disordered phase
of $\phi$, after integrating out the massive $\phi$ and dynamical
gauge field $A_\mu$, the lowest order contribution to the
effective Lagrangian of $A^{ext}_\mu$ is still a Maxwell term: $
\mathcal{L}_{eff} \sim (\partial A^{ext})^2 + c \theta (\partial^2
A^{ext})(\partial A^{ext}) + \cdots$. While in the condensate of
$\phi$, the effective Lagrangian of $A^{ext}_\mu$ acquires a
Chern-Simons term at level 1. This analysis implies that this
transition is equivalent to a topological transition between a
trivial insulator and a Chern insulator with Chern number 1. The
universality class of this type of topological transition of Chern
insulator is very well-understood, it can be simply described by a
2+1d Dirac fermion: \beqn \mathcal{L} = \bar{\psi}\gamma_\mu
\partial_\mu \psi + m \bar{\psi}\psi, \eeqn here the trivial
insulator and Chern insulator correspond to $m > 0$ and $m < 0$
respectively, and $m = 0$ corresponds to the quantum critical
point at $r = 0$ in Eq.~\ref{ntheta} with $N = 1$. Thus we
conjecture that when $N = 1$ and $\theta = \pi/2$, the critical
point in Eq.~\ref{ntheta} is dual to a massless free Dirac
fermion. In Ref.~\cite{fisherwu} a similar conjecture was made
that the 3D XY transition is dual to a massless Dirac fermion
coupled to a noncompact U(1) gauge field.

\section{Summary and Discussion}

In this work we have discussed a possible theory for the direct
continuous transition between the $Z_2$ liquid phase and the
N\'{e}el order, and this is a candidate theory for the
liquid-N\'{e}el transition observed in Ref.~\cite{meng,hongchen}.
We have taken the square lattice as an example, but results
discussed in this paper can also be applied to the honeycomb
lattice after straightforward generalization.

In our theory, we exploited the fact that in two dimension, the
$e$ and $m-$type excitations are both point like defects, thus
their nontrivial statistics can be described well with a mutual
Chern-Simons theory. By contrast, in a three dimensional $Z_2$
liquid phase, there is a mutual semion statistics between the
point particle like $e-$excitation and loop like $m-$type
excitation. Thus the effective field theory for the three
dimensional $Z_2$ liquid phase is the so-called BF theory
$\mathcal{L}_{eff} \sim \frac{i}{\pi} \epsilon_{\mu\nu\rho\tau}
a_\mu \partial_\nu b_{\rho\tau}$, where $a_\mu$ is the U(1) gauge
field that couples to the $e-$type point particle, and
$b_{\mu\nu}$ is an antisymmetric rank-2 antisymmetric tensor gauge
field that couples to the $m-$type loop excitation. The global
phase diagram around the three dimensional $Z_2$ liquid phase is
another interesting subject, and we will leave it to future
studies.
%EM ``is`` was missing.

%The main evidence for the existence of the $Z_2$ liquid phase is
%the topological entanglement entropy: $\beta = \ln 2$. Based on
%this entanglement entropy, there is another possible and more
%exotic candidate state: the time-reversal symmetry breaking
%SU(2)$_2$ nonabelian topological state, which has the same
%topological entanglement entropy as the $Z_2$ spin liquid
%\cite{kitaeventropy,wenentropy}. The difference between the
%SU(2)$_2$ and the $Z_2$ spin liquid phase can in principle be
%detected through measuring the correlation of the time-reversal
%symmetry breaking spin chirality operator $\vec{S}_i \cdot
%(\vec{S}_j \times \vec{S}_k)$, which should vanish in the $Z_2$
%topological state, but have a nonzero expectation value in the
%SU(2)$_2$ state. These two states also have very different
%entanglement spectrum, as the edge states of SU(2)$_2$ topological
%phase is a $c = 3/2$ conformal field theory, while there is no
%nontrivial edge states for the $Z_2$ liquid phase.

%The quantum phase transition between the SU(2)$_2$ state and the
%ordinary N\'{e}el state is also an interesting subject, we will
%leave this problem to future studies.

\begin{acknowledgements}

This work is supported by the Sloan Foundation. The authors thank
Matthew Fisher, Zhengcheng Gu, Chetan Nayak, Subir Sachdev,
Senthil Todadri, and Xiao-Gang Wen for very helpful discussions.

\end{acknowledgements}

%\bibliography{triangle}

\begin{thebibliography}{0}

\bibitem{white} Simeng~Yan, David~A.~Huse, Steven~R.~White, {\bf 332}, 1173, (2011).

\bibitem{jiang2008} H.~C.~Jiang, Z.~Y.~Weng, D.~N.~Sheng, Phys. Rev. Lett. {\bf 101}, 117203 (2008).

\bibitem{hongchen} Hong-Chen~Jiang, Hong~Yao, Leon~Balents, arXiv:1112.2241, (2011).

\bibitem{gu2011} Ling~Wang, Zheng-Cheng~Gu, Frank~Verstraete, Xiao-Gang~Wen, arXiv:1112.3331, (2011).

\bibitem{fabio} Fabio Mezzacapo, arXiv:1203.6381, (2012).

\bibitem{balentsz2} L.~Balents, M.~P.~A.~Fisher, and S.~M.~Girvin, {\it Phys. Rev. B} {\bf 65}, 224412 (2002).

\bibitem{kimz2} Sergei.~V.~Isakov, Yong-Baek~Kim, and A.~Paramekanti, {\it Phys. Rev. Lett.} {\bf 97}, 207204 (2006).

\bibitem{kimz22} Sergei.~V.~Isakov, A.~Paramekanti, and Yong-Baek~Kim,  {\it Phys. Rev. B} {\bf 76}, 224431 (2007).

\bibitem{melko2} Sergei~V.~Isakov, Matthew~B.~Hastings, Roger~G.~Melko, {\it Nature Physics} {\bf 7}, 772 (2011).

\bibitem{melko3} Sergei~V.~Isakov, Roger~G.~Melko, Matthew~B.~Hastings, {\it Science} {\bf 335 }, 193 (2012).

\bibitem{meng} Z.~Y.~Meng, T.~C.~Lang, S.~Wessel, F.~F.~Assaad, A.~Muramatsu, {\it Nature} {\bf 464}, 847 (2010).

\bibitem{kitaev1997} A.~Kitaev, Annals Phys., {\bf 303},2 (2003).

\bibitem{senthilchubukov} Andrey~V.~Chubukov, Subir~Sachdev, T.~Senthil, {\it Nucl. Phys. B} {\bf 426}, 601 (1994).

\bibitem{ngs} Satoru~Nakatsuji, Yusuke~Nambu, Hiroshi~Tonomura, Osamu~Sakai,
Seth~Jonas, Collin~Broholm, Hirokazu~Tsunetsugu, Yiming~Qiu and
Yoshiteru~Maeno, {\it Science} {\bf 309}, 1697 (2006).

\bibitem{wang} Fa~Wang, {\it Phys. Rev. B} {\bf 82}, 024419 (2011).

\bibitem{ran} Yuan-Ming~Lu, Ying~Ran, {\it Phys. Rev. B} {\bf 84}, 024420 (2011).

\bibitem{xu2010} Cenke~Xu, Phys. Rev. B {\bf 83}, 024408 (2011).

\bibitem{xuqi} Yang~Qi, Cenke~Xu, Subir~Sachdev, {\it Phys. Rev. Lett.} {\bf 102}, 176401 (2009).

\bibitem{wen} Xiao-Gang Wen and Y.-S. Wu, Phys. Rev. Lett. {\bf 70}, 1501 (1993).

\bibitem{xusachdev} Cenke~Xu, Subir~Sachdev, Phys. Rev. B {\bf 79}, 064405 (2009).

\bibitem{xubalents} Cenke~Xu, and Leon~Balents, {\it Phys. Rev. B} {\bf 84}, 014402 (2011).

\bibitem{kou} Su-Peng~Kou, Michael~Levin, and Xiao-Gang~Wen, {\it Phys. Rev. B} {\bf 78}, 155134 (2008).

\bibitem{deconfine1} T.~Senthil, Ashvin~Vishwanath, Leon~Balents, Subir~Sachdev, M.~P.~A.~Fisher, {\it Science} {\bf 303}, 1490 (2004).

\bibitem{deconfine2} T.~Senthil, Leon~Balents, Subir~Sachdev, Ashvin~Vishwanath, Matthew~P.~A.~Fisher, {\it Phys. Rev. B} {\bf 70}, 144407 (2004).

\bibitem{sachdevcpn} Ribhu~K.~Kaul, and Subir~Sachdev, {\it Phys. Rev. B} {\bf 77}, 155105 (2008).

\bibitem{sandvik1} Anders W. Sandvik, {\it Phys. Rev. Lett.} {\bf 98}, 227202 (2007).

\bibitem{sandvikvbs} Anders W. Sandvik, {\it Phys. Rev. B} {\bf 85}, 134407 (2012).

\bibitem{ribhu} Roger G. Melko and Ribhu K. Kaul, {\it Phys. Rev. Lett.} {\bf 100}, 017203 (2008).

\bibitem{swinglesenthil} Brian~Swingle, and T.~Senthil, arXiv:1109.3185.

\bibitem{sachdeventropy} Igor~R.~Klebanov, Silviu~S.~Pufu, Subir~Sachdev, Benjamin~R.~Safdi, arXiv:1112.5342.

\bibitem{fisherwu} Wei~Chen, Matthew~P.~A.~Fisher, and Yong-Shi~Wu, Phys. Rev. B {\bf 48}, 13749 (1993) .

%\bibitem{kitaeventropy} Alexei~Kitaev and John~Preskill, {\it Phys. Rev. Lett.} {\bf 96}, 110404 (2006).

%\bibitem{wenentropy} Michael~Levin and Xiao-Gang~Wsen, {\it Phys. Rev. Lett.} {\bf 96}, 110405 (2006).

\end{thebibliography}

\end{document}